\documentclass[twocolumn]{aastex6}

\collaborationName{Friends of AASTeX}

\begin{document}


\title{Are there multiple populations of Fast Radio Bursts?}


\author{Divya Palaniswamy, Ye Li, and Bing Zhang }
\affil{Department of Physics and Astronomy, University of Nevada, Las Vegas, Las Vegas, NV 89154,
zhang@physics.unlv.edu}

\begin{abstract}
The repeating FRB 121102 (the ``repeater'') shows {repetitive bursting activities} and was localized in a host galaxy at $z=0.193$. On the other hand, despite {dozens} of hours of  telescope time spent on follow-up observations, no other FRBs have been observed to repeat. Yet, it has been speculated that the repeater is the {prototype} of FRBs, and that other FRBs should show similar repeating patterns. Using the published data, we compare the repeater with other FRBs in the \textit{observed} time interval ($\Delta t$) - flux ratio ($S_i / S_{i+1}$) plane.  We find that whereas other FRBs occupy the upper (large $S_i / S_{i+1}$) and right (large $\Delta t$) regions of the plane due to the non-detections of other bursts, some of the repeater bursts fall into the lower-left region of the plot (short interval and small flux ratio) excluded by the non-detection data of other FRBs. The trend also exists even if one only selects those bursts detectable by the Parkes radio telescope. If other FRBs were similar to the repeater, our simulations suggest that the probability that none of them have been detected to repeat with the current searches would be $\sim (10^{-4}-10^{-3})$. We suggest that the repeater is not representative of the entire FRB population, and that there is strong evidence of more than one population of FRBs.
\end{abstract}

\section{Introduction} \label{sec:intro}

A decade ago a mysterious new class of millisecond-duration radio transients called Fast Radio Bursts (FRBs) were discovered \citep{lorimer2007}. These are highly dispersed, bright ($\rm \sim Jy$) {bursts detected from high galactic latitudes} \citep{lorimer2007,Keane2012, thor2013,Spitler2014, Pet2015, Masui2015, Keane2016, cham2016, Ravi2016, Bannister2017, Petroff2017}. Since the first discovery, twenty-six FRBs have been discovered and twenty two have been published\footnote{http://www.astronomy.swin.edu.au/pulsar/frbcat/}. The large dispersion measure (DM), if interpreted as due to dispersion of free electrons in the intergalactic medium, implies that the sources are at cosmological distances, e.g. in the redshift range $0.2-1.5$. The isotropic energy release of FRBs is estimated to be in the range of $\sim 10^{40-41}$ erg \citep[e.g.][]{thor2013}.

Out of these detected FRBs, one source, FRB 121102 (also known as the ``repeater''), clearly repeats \citep{Spitler2016, Scholz2016}. Thanks to its repeating behavior, {a compact, steady radio counterpart associated with the bursting source} was detected  \citep{Chatterjee2017,Marcote2017} and a star formation host galaxy at $z=0.193$ was identified \citep{Tendulkar2017}, which provides solid evidence of the cosmological origin of at least this FRB source. {Most other FRBs have been re-observed after their discoveries}, but so far no positive detection of a repeating burst has been reported \citep[e.g.][]{Petroff2015,cham2016}. One possible exception might be the FRB 110220 / FRB 140514 pair, which are in the same 14.4 arcmin beam and are 9 arcmin apart. \cite{Piro2017} hypothesized that they may originate from the same neutron star embedded within a supernova remnant (SNR) that provides an evolving DM as the ejecta expands. A detection of a third burst with an even smaller DM from this location  would support this hypothesis. In any case, {it was speculated by some authors  \citep[e.g.][]{Spitler2016,Lu2016}} that other FRBs are not much different from the repeater, and the non-detection of the repeating bursts from those sources might be due to the lower sensitivity of the Parkes telescope ({which detected most of the FRBs published so far}) than the Arecibo 300-m telescope (which detected many bursts from the repeater). On the other hand, it has been also suggested \citep{Keane2016} that there might be more than one class of FRBs: e.g. repeating and non-repeating ones. The non-repeating FRBs are found to be usually unresolved, whereas the repeating bursts of FRB 121102 are all resolved with a temporal structure.

Many FRB progenitor models have been proposed in the literature. Some models invoke non-catastrophic events such as giant magnetar flares \citep{pop2010,kulkarni14,katz16}, giant pulses from young pulsars \citep{Cordes2016, Connor2016} and young rapidly-spinning magnetars \citep{Metzger2017,Kashi2017}, repeated captures of asteroids by a neutron star \citep{Dai2016}, and repeated ``cosmic combing'' events when an astrophysical stream interacts with the magnetosphere of an old neutron star \citep{Zhang2017}. Alternatively, other classes of models invoke catastrophic events, such as the collapse of supra-massive neutron stars (the so-called ``blitzars") \citep{FR2014, Zhang2014} and mergers of compact stars (NS-NS, NS-BH, BH-BH) \citep{Tot2013, Zhang2016, wang16}. The FRBs produced from these systems would not repeat.

In this paper, we use the published detection data of the repeating FRB 121102 and the non-detection data of other FRBs to investigate whether the FRB 121102 source is representative of all FRBs in terms of its repeating behavior. 

\section{FRB Repeating time interval - Flux Ratio Distribution}

If other FRBs also repeat but no repeating bursts are detected, {it could be due to either of the following two reasons}: (1) The repeating time interval (duty cycle) is long, and the current observation time is not long enough to catch another burst yet; (2) {Other repeating bursts might have occurred, but they are too faint to be detected by the searching telescopes (e.g. Parkes).} In order to quantify these two effects, we introduce two parameters: the {\em observed} time interval between bursts, $\Delta t$, and the peak flux ratio between two adjacent bursts, $S_{i}/S_{i+1}$. The former addresses the duty cycle of the repeating bursts, whereas the latter addresses the probability that the additional bursts are too faint to be detected.

\subsection{The Raw $\Delta t - (S_i/S_{i+1})$ Plot}

\begin{figure}
	\epsscale{1}
	\plotone{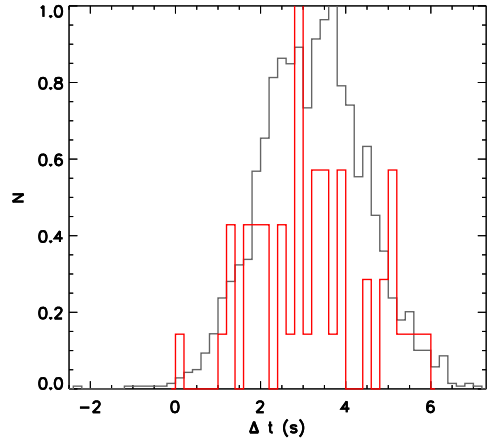}
	\caption{The simulated $\Delta t$ distribution (red) compared with the input $\Delta t_{\rm intr}$ distribution with a log-Gaussian distribution.}
	\label{fig:Delta-t}
\end{figure}

For both repeating FRB 121102 and other non-repeating FRBs, observations were initially carried out in a certain time interval. Later the sources were re-observed multiple times. For each source, there are several time gaps during which no telescope was observing the source. As a result, it is impossible to precisely define the intrinsic time intervals between two adjacent bursts. In our analysis, we define an {\em observed} time interval $\Delta t$ as the time interval between bursts during the observational time span only, so that the time gaps during which no observation was carried out are removed. In order to test the validity of using $\Delta t$, we perform a set of Monte Carlo simulations by assuming different (uniform, power law, Gaussian, and log-Gaussian) distributions of the intrinsic time interval $\Delta t_{\rm intr}$ between adjacent bursts. By randomly choosing observational time intervals following a uniform distribution, we find that regardless of the $\Delta t_{\rm intr}$ distribution, the observed $\Delta t$ distribution essentially follows the $\Delta t_{\rm intr}$ distribution if the number of observational windows is large enough. As an example, we show in Fig.\ref{fig:Delta-t} the simulated $\Delta t$ distribution (red) against a log-Gaussian $\Delta t_{\rm intr}$ distribution (black), with the number of observations comparable to that for FRB 121102.  We therefore use the $\Delta t$ data of FRB 121102 to derive the $\Delta t_{\rm intr}$ distribution of FRBs in general (assuming that they are similar to each other).

For FRB 121102, 
the first burst was discovered in the PALFA survey at 1.4 GHz \citep{Spitler2014}. Later an extensive follow-up observation using Arecibo resulted in the detection of 10 more bursts \citep{Spitler2016}. Another six bursts from this source were later detected and published: five from Green Bank telescope (GBT) at 2.0 GHz and one at 1.4 GHz with the Arecibo Observatory (AO). Later, 14 more bursts from FRB121102 were detected and published: ten from GBT at 2.0 GHz and four from AO at 1.4 GHz \citep{Scholz2017}. In our analysis, we include a total of 40 bursts based on the published data, even though by now $> 140$ bursts from this source have been detected, but are not published yet (L. Spitler, 2017, talk presented at the Aspen FRB conference, {http://aspen17.phys.wvu.edu/Spitler.pdf}). Including all $>140$ bursts in our analysis will only strengthen the conclusion of this paper. 

\begin{figure}
	\epsscale{1}
	\plotone{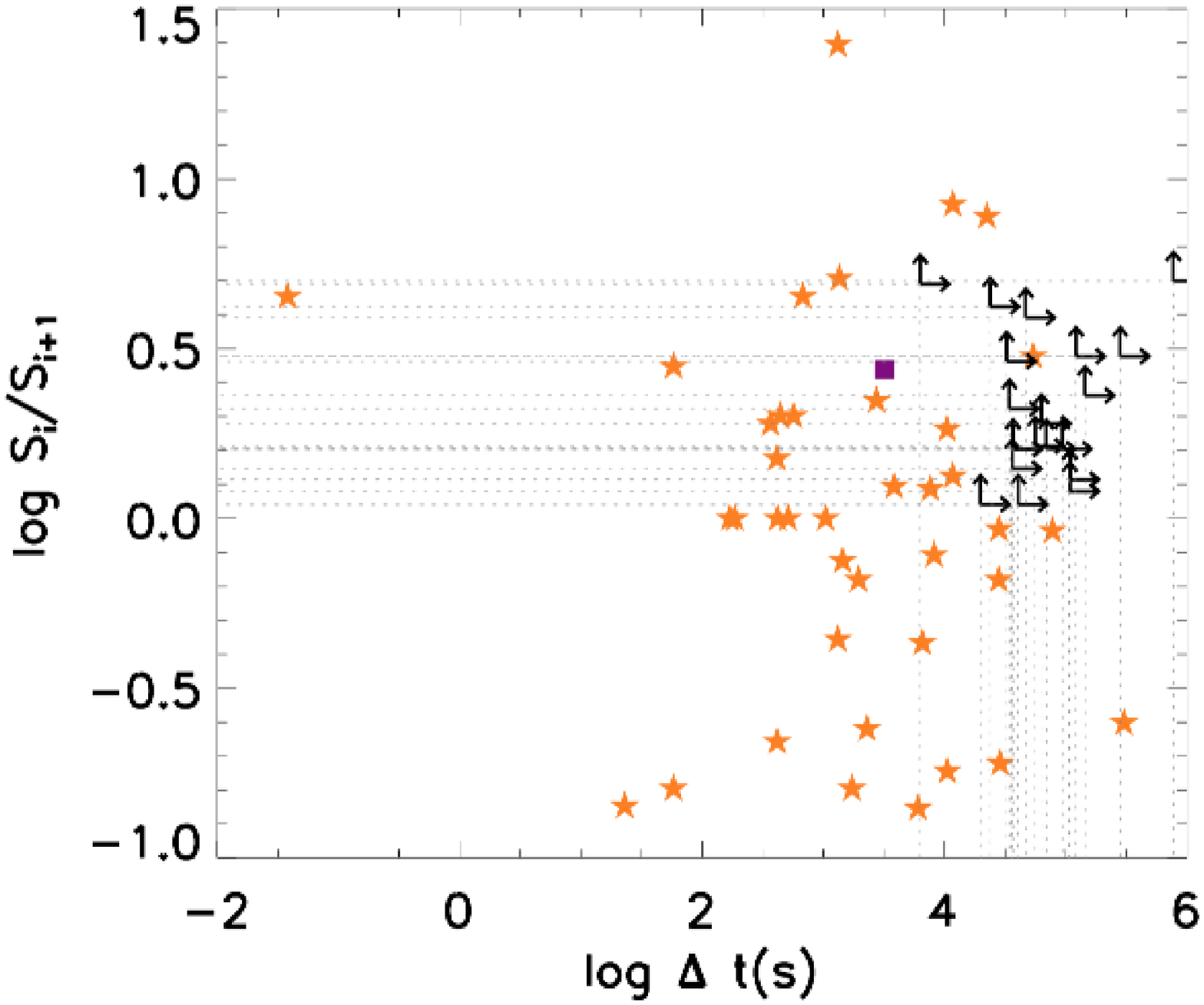}
	\plotone{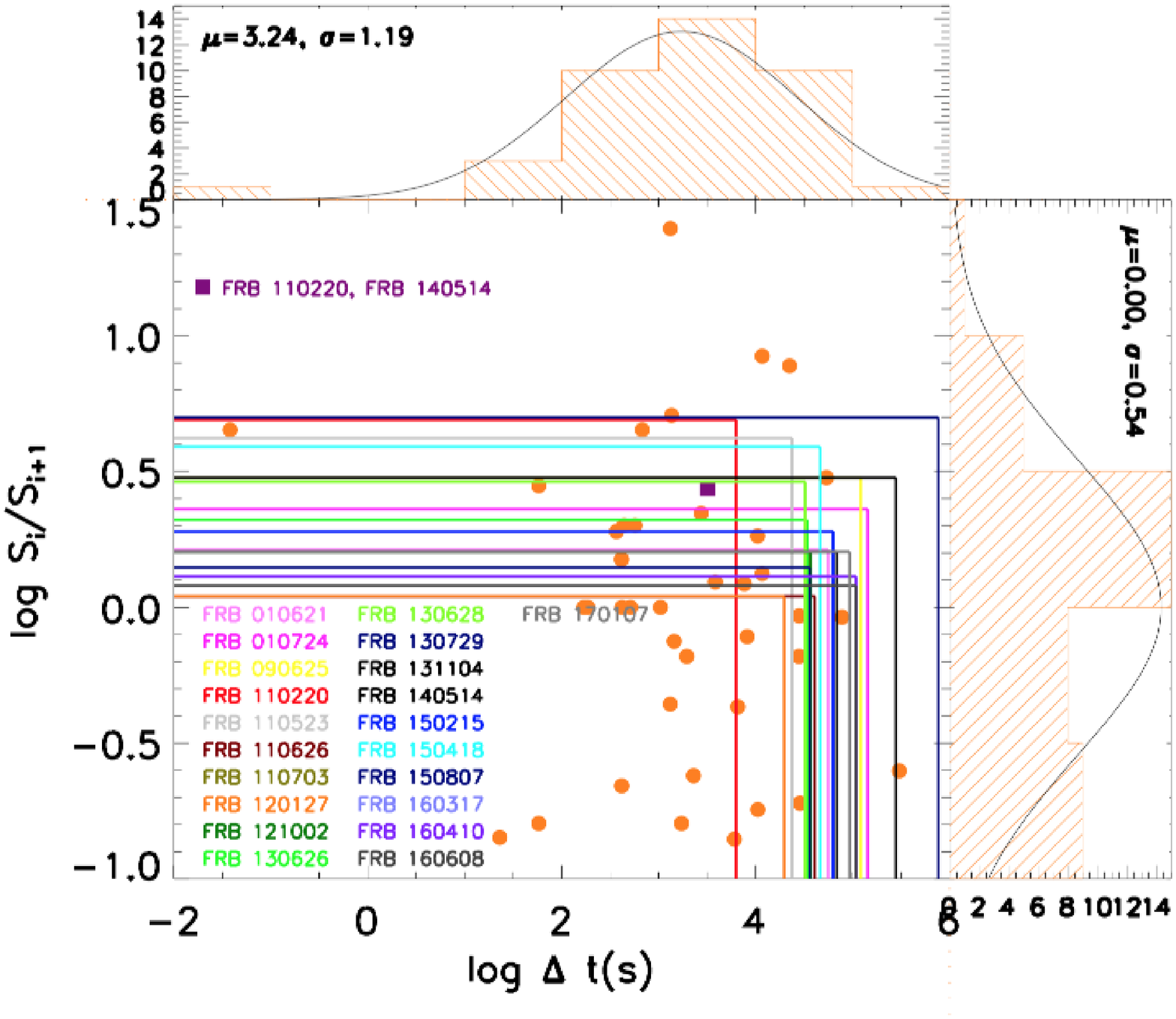}
	\caption{(a) The raw $\Delta t$ - $S_{i}/S_{i+1}$ distribution plot of the repeater (orange stars) and other non-repeating FRBs (black points with right and upper arrows). For other FRBs, based on non-detections, only lower limits of $\Delta t$ and $S_i/S_{i+1}$ are presented, which are indicated as the regions above the horizontal dashed line or to the right of the vertical dashed line of each FRB. FRB110220 and FRB140514 are both presented separately (assuming that they are related) and jointly (assuming they are from the same source, purple square). (b) The same plot but with individual FRBs identified. The histograms of $\log \Delta t$ and $\log (S_i/S_{i+1})$ distributions are plotted along with the best-fit distributions that are used in later Monte Carlo simulations.}
	\label{fig:FRB_TS1}
\end{figure}

Bursts from this source are irregular and clustered in time 
\citep{Scholz2016}. The total telescope time spent on FRB121102 was over 177 hours for the observations of the 40 bursts. Among the 40 bursts, six bursts were found within a 10 min of observing period, and four were detected in a 20-min observing window \citep{Scholz2016}. Recently, \cite{Scholz2017} reported two bursts that are 0.038 seconds apart. The total observing time is spread over telescopes with different sensitivities, and the observations were done in a range of radio frequencies. For completeness, we include all the bursts regardless of the observing telescope and frequency\footnote{Ignoring repeating bursts detected by telescopes other than Arecibo would introduce a bias of the repeating duty cycle of the source. It would be more appropriate to include all the available information in this analysis, even though the data are not uniform. An attempt of removing the non-uniformity of the data is presented below in Section \ref{sec:Parkes}.}. The observed time intervals $\Delta t$ are calculated for each pair of adjacent bursts defined by the sequence numbers $i$ and $i+1$\footnote{The definition of the burst number is arbitrary. Given two adjacent bursts defined, one can derive $S_i/S_{i+1}$ and $\Delta t$. When a higher detection threshold is defined (e.g. for Parkes-threshold bursts only, see Sect. \ref{sec:Parkes}), the burst numbers are re-ordered, and $S_i/S_{i+1}$ and $\Delta t$ are re-defined.}. Table \ref{tab:observing time_repeator} presents $\Delta t$ of all adjacent pairs of the bursts, which are also presented in Figure \ref{fig:FRB_TS1}.

Other FRBs (detected by Parkes and GBT) have been extensively followed up on a variety range of timescales, {most with dozens of hours of observing times (Table \ref{tab:observing time} and references therein)}. No additional bursts have been detected. We include all the observing times {\em after} the detection of the original burst and define it as the lower limit of $\Delta t$ for the bursts whose flux reaches the flux threshold. The data are shown in Table \ref{tab:observing time} and presented in Figure \ref{fig:FRB_TS1}.

The flux ratio is defined as $S_i/S_{i+1}$, where $S_{i}$ and $S_{i+1}$ are the peak fluxes of $i$-th and $(i+1)$-th burst. It is greater/less than one if the next FRB is fainter/brighter.  The peak fluxes and pulse widths for each burst from FRB 121102 are different. For completeness, we include all the bursts to define $ S_{i}/S_{i+1}$, most of which were detected at 1.4 GHz, but some of which were detected at 2 GHz. They are presented in Figure \ref{fig:FRB_TS1}.

For other FRBs, since no repeating bursts have been detected, any possible burst during the observation time frame, if any, must have a peak flux that is faint enough to evade detection.  {For these FRBs, we can only plot the lower limit of $S_i/S_{i+1}$, which is simply $S_1/S_{th}$, where $S_{th}$ is the $10\sigma$ detection flux with respect the receiver noise level. Such a threshold has been widely adopted to claim the detection of an FRB. The values of $S_1/S_{th}$ are adopted as the lower limit of $S_i/S_{i+1}$ as shown in Table \ref{tab:observing time}.}

The raw $\Delta t - S_{i}/S_{i+1}$ plot is shown in Figure \ref{fig:FRB_TS1}a. The orange stars denote the repeating bursts from FRB 121102, while for other bursts, the allowed regions are above the horizontal line and to the right of the vertical line for each burst (as denoted by the right and upper arrows). One can immediately see that some bursts from the repeating FRB 121102 are located in the region (lower left region) in the $\Delta t - S_{i}/S_{i+1}$ plane that is excluded by the data of other FRBs. Compared with other FRBs, for the repeater the time intervals between bursts are much shorter, and the flux ratios are smaller, some of them being even less than unity (suggesting that the later burst is brighter than the earlier one). 

Assuming that FRB 110220 and FRB 140514 belong to the same source, as argued by \cite{Piro2017}, we plot another point (purple square) based on the detected values. This point sits between the region of FRB 121102 and the region of other FRBs. On the other hand, the non-detection of other bursts following FRB 140514 again places the second point of the source to the region of other FRBs. 

\begin{deluxetable}{cccccc}\label{tab:table1}
		\tablecaption{\small{The observational information of the 40 bursts detected from the repeater FRB 121102. Column 1 gives the burst number, column 2 gives the receiver and telescope name, column 3 gives the observed time interval $\Delta t$ between the burst $i$ and $i+1$, column 4 gives the peak flux density for each burst, column 5 gives the flux ratio $S_i/S_{i+1}$, and column 6 gives the references.} \label{tab:observing time_repeator}}
		\tabletypesize{\scriptsize}
	\tablehead{
		\colhead{Burst } &
		\colhead{Telescope/}  & \colhead{$\Delta t $} &  \colhead{Peak } & Flux Ratio &  Ref\\
		\colhead{Number} & \colhead{Receiver} &   \colhead{(s)} & \colhead{ Flux (Jy)}  & \colhead{($S_{i}/S_{i+1}$)} 
	}
	\startdata
	1		&	AO/ALFA	&	128		&		0.04	&		0.00 & [1,2] \\			
	2		&	AO/ALFA	&	11694	&		0.03	&		1.33 & [2,3]\\
	3		&	AO/ALFA	&	512		&		0.03	&		1.00 & [2,3]\\
	4		&	AO/ALFA	&	1444	&		0.04	&		0.75 & [2,3]\\
	5		&	AO/ALFA	&	569		&		0.02	&		2.00 & [2,3]\\
	6		&	AO/ALFA	&	419		&		0.02	&		1.00 & [2,3]\\
	7		&	AO/ALFA	&	23		&		0.14	&		0.142 & [2,3]\\
	8		&	AO/ALFA	&	58		&		0.05	&		2.80  & [2,3]\\
	9		&	AO/ALFA	&	186		&		0.05	&		1.00 & [2,3]\\
	10		&	AO/ALFA	&	169		&		0.05	&		1.00 & [2,3]\\	
	11		&	AO/ALFA	&	58		&		0.31	&		0.16 & [2,3]\\
	12		&	GBT/S-band	&22400	&		0.04	&		7.75  & [3]\\
	13		&	GBT/S-band	&28033	&		0.06	&		0.66 & [3]\\
	14		&	GBT/S-band	&414	&		0.04	&		1.50 & [3]\\
	15		&	GBT/S-band	&442	&		0.02	&		2.00  & [3]\\
	16		&	GBT/S-band	&416	&		0.09	&		0.22 & [3]\\
	17		&	AO/L-WIDE	&53871	&		0.03	&		3.00  & [2,3]\\
	18      &   VLA/3GHz    &302382  &      0.12    &        0.25  & [4,5]\\
	19      &    VLA/3GHz   & 10543  &      0.67    &        0.18  & [4,5] \\
	20      &    VLA/3GHz   & 1321   &      0.027   &        24.81 & [4,5] \\
	21      &    VLA/3GHz   & 6556   &      0.063   &        0.43  & [4,5] \\
	22      &    VLA/3GHz   & 28736  &      0.328   &        0.19  & [4,5]\\
	23      &    VLA/3GHz   & 11701  &      0.039   &        8.41  & [4,5] \\
	24      &    VLA/3GHz   & 8185   &      0.050   &        0.78  & [4,5] \\
	25      &    GBT/S-band & 6049   &      0.36    &        0.14  & [6] \\
	26      &    GBT/S-band  &0.038  &         0.08  &          4.5 & [6]  \\
	27      &    VLA/3GHz    &28143  &         0.086 &         0.93  & [4,5]\\
	28      &    GBT/S-band    &2292   &         0.36  &          0.24 & [6]  \\
	29      &    GBT/S-band    &3836   &         0.29  &          1.24  & [6] \\
	30      &    VLA/3GHz    &10500  &         0.158 &          1.83   & [6] \\
	31      &    GBT/S-band    &77653  &         0.17  &          0.92  & [6]  \\
	32      &    GBT/S-band    &1319   &         0.38  &          0.44  & [6]  \\
	33      &    GBT/S-band    &367    &        0.20   &         1.9    & [6]  \\
	34      &    GBT/S-band    &2745   &         0.09  &          2.22  & [6]  \\
	35      &    GBT/S-band    &1723   &         0.56  &          0.16   & [6]\\
	36      &    GBT/S-band    &1358   &         0.11  &          5.09   & [6] \\
	37      &    AO/1.4GHz     &7636   &         0.09  &          1.22   & [6] \\
	38      &    AO/1.4GHz      &678    &         0.02  &          4.5    & [6] \\
	39      &    AO/1.4GHz      &1045   &         0.02  &          1.0    & [6] \\
	40      &    AO/1.4GHz      &1949   &         0.03   &         0.66   & [6] \\	
	\enddata
	\tablecomments{[1] \cite{Spitler2014}; [2] \cite{Spitler2016}; [3] \cite{Scholz2016}; [4] \cite{Chatterjee2017}; [5] \cite{Law2017}: [6] \cite{Scholz2017}}
\end{deluxetable}

\begin{deluxetable}{cccc}\label{tab:table2}
        \tablecaption{The follow-up observation information of other FRBs that do not show repeating bursts, including the total number of observation hours and the lower limit of $S_i/S_{i+1}$, which is defined by $S_1/S_{th}$.
\label{tab:observing time}}
        \tablehead{
                \colhead{FRB Name} &
                \colhead{Total hours} & \colhead{Flux Ratio} & \colhead{Reference} \\
                \colhead{(yymmdd)} & \colhead{(hours)} & \colhead{$S_{i}/S_{i+1}$} 
        }

        \startdata
FRB010621 &  15.5  & $>$  1.63 &  [6,11]  \\
FRB010724 &  40  & $>$  2.30 &  [7]  \\
FRB090625 &  33.65  & $>$  3.00 & [1,3]  \\
FRB110220 &  1.75  & $>$  4.90 &  [1,2]  \\
FRB110523 &  6.6  & $>$  4.20 &  [15]  \\
FRB110626 &  11.25  & $>$  1.10 &  [1,2]  \\
FRB110703 &  10.1  & $>$  1.60 &  [1,2]  \\
FRB120127 &  5.5  & $>$  1.10 &  [1,2]  \\
FRB121002 &  10.25  & $>$  1.60 &  [1,3]  \\
FRB130626 &  9.5  & $>$  2.10 & [1,3]  \\
FRB130628 &  9.0  & $>$  2.90 &  [1,3]  \\
FRB130729 &  10  & $>$  1.40 &  [1,3]  \\
FRB131104 &  78  & $>$  3.00 &  [8]  \\
FRB140514 &  19.2  & $>$  1.60 &  [4]  \\
FRB150215 &  17.5  & $>$  1.90 &  [14]  \\
FRB150418 &  13.0  & $>$  3.90 &  [9]  \\
FRB150807 &  215  & $>$  5.00 & [10]  \\
FRB160317 &  30.0  & $>$  1.30 & [12]  \\
FRB160410 &  30.0  & $>$  1.30 & [12]  \\
FRB160608 &  30.0  & $>$  1.20 & [12]  \\
FRB170107 &  26.1  & $>$  1.60 &  [13]  \\
        \enddata
        \tablenotetext{}{$\dagger$ - Follow up hours are not published}
        \tablecomments{[1]\cite{Petroff2015}; [2]\cite{thor2013} [3] \cite{cham2016} [4] \cite{Pet2015} [5] \cite{Bruke2014} [6] \cite{Keane2012} [7]\cite{lorimer2007} [8] \cite{Ravi2015} [9] \cite{Keane2016} [10]\cite{Ravi2016}  [11] \cite{Bannister2014} [12] \cite{Caleb2017} [13] \cite{Bannister2017} [14] \cite{Petroff2017}}
\end{deluxetable}

\subsection{Correcting to the sensitivity limit of Parkes}\label{sec:Parkes}

Since Arecibo is more sensitive than Parkes, a more rigorous treatment of the repeating FRB 121102 should include only those bursts that are above the Parkes sensitivity threshold for a fair comparison with the results of other FRBs. 

\begin{figure}
	\epsscale{1}
	\plotone{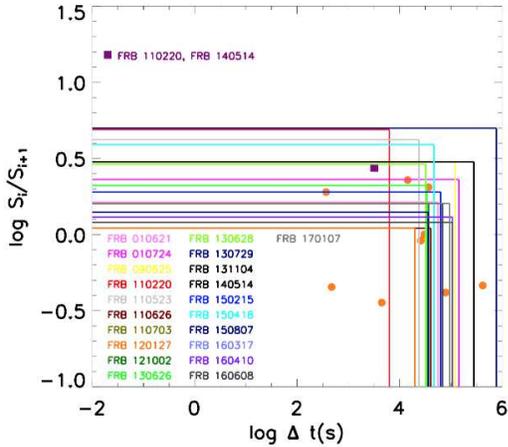}
	\caption{The $\Delta t - S_i/S_{i+1}$  plot after only selecting bursts detectable with the Parkes telescope and at 1.4 GHz. The repeater has ten bursts above the Parkes threshold.} 
	\label{fig:ts2}
\end{figure}

The peak flux densities (for pulse widths ranging from $1.28$ - $8.192$ ms using a signal-to-noise ratio of 10.0) of Parkes FRBs are in the range  $0.11$ - $0.28$ Jy. Out of the 40 repeating bursts from FRB 121102, ten bursts (burst numbers 7, 11, 19, 22, 25, 28, 30, 32, 33, 35) would have been above the Parkes S/N cutoff limit at 1.4 GHz. 
For example, both bursts 7 and 11 were originally detected by Arecibo within a 10 min observation window \citep{Spitler2016}. 
Bursts 8, 9, and 10 would not be detected (for a flat spectrum) by Parkes, we therefore pretend that they were not detected and redefine $\Delta t$ and $S_i/S_{i+1}$ and plot the points in Figure \ref{fig:ts2}. Out of the 15 bursts detected by GBT at 2 GHz, five of them (number 25, 28, 32, 33, 35) would have been detected by Parkes for a flat spectrum\footnote{Burst numbers 19, 22, 33 where detected by Very Large Array at 3 GHz, and would be also detected by Parkes for a flat spectrum. The spectral indices of the repeating bursts of FRB 121102 vary significantly from burst to burst \citep{Spitler2016}, so it is hard to reliably extrapolate the fluxes from 2GHz (GBT) and 3 GHz (VLA) to 1.4 GHz. In any case, for indicative purposes we assume a flat spectrum for all the bursts and include all the bursts detected by GBT and VLA in our analysis. Introducing more precise spectral indices would modify some data points presented in Figs. \ref{fig:FRB_TS1} and \ref{fig:ts2}, but would not change the conclusion of this paper.}. The results are presented in Figure \ref{fig:ts2}.



Ten bursts can define nine points in Figure \ref{fig:ts2}. One can see that although most of the points still fall in the regime allowed by the non-detection limits of other FRBs, two points lie in the lower-left corner of the plot, a region excluded by the non-detection data of other FRBs. Since the total observing hours of other FRBs are usually long (two of them, FRB131104 and FRB150807, each has follow-up hours more than 70 hours), at least some repeating bursts should have been detected if all other FRBs are similar to FRB 121102. The non-detection of any repeating bursts from any of those FRBs therefore suggests that FRB 121102 is likely not representative of the FRB population.

\subsection{Monte Carlo simulations}

In order to quantify the probability of non-detection any burst if all other FRBs are similar to FRB 121102,
we perform Monte Carlo simulations of typical repeating FRBs based on the $\Delta t$ and 
$S_i/S_{i+1}$ distributions as observed in FRB 121102.
The distributions are presented as the histograms in the upper and right panels of Figure \ref{fig:FRB_TS1}b.
They are fitted with logarithmic Gaussian distributions with Cash statistics \citep{Cash1979}.
The best fits are presented as solid curves overlapping with the histograms, with median values and standard deviations labelled. 

There were hundreds of pointings on FRB 121102.
The intervals between these pointings (in units of seconds) generally follow a logarithmic Gaussian distribution,
with a median value 4.9 and standard deviation 0.9.
Each pointing has a duration from hundreds to tens of thousands of seconds,
with the logarithmic median value 3.6 and standard deviation 0.4.
Following these realistic observational properties,
we simulate 500 pointings with logarithmic Gaussian distributed durations and intervals.
Tens of bursts are ``detected'',
and their observed time intervals follow what was observed.

In order to test whether other non-repeating FRBs are similar to FRB 121102, we simulate other FRBs with the same 
$\Delta t$ and $S_i/S_{i+1}$ distributions as FRB 121102.
Non-repeating FRBs were typically observed for tens of hours.
We simulate the observational intervals and durations of these FRBs similar to FRB 121102, and keep the 
total observational time as the true observational time of each FRB. We notice that the flux range of FRB 121102 is by a factor $0.67/0.02 \sim 34$. We therefore require that the simulated fluxes of other bursts from the non-repeating FRBs have a flux distribution by a factor of 40. One important parameter to perform the simulations is the normalization of the flux distribution. We consider two cases for the flux distribution: {\em Case I:} The flux range is (0.05 - 2) times of the detected flux. This assumes that the detected burst belongs to one of the brightest among the bursts, which is possible due to the observational selection effect (brighter ones are easier to detect). Notice that this is the most conservative case in terms of detecting repeating bursts, since most  other bursts are fainter than this one and more likely to be below the sensitivity threshold. {\em Case II:} This possibility assumes that the detected burst is a  normal one, so that the flux range is (0.16 - 6.3) times of the detected flux.

Based on the detected flux of each FRB, we randomly choose $S_i/S_{i+1}$ and $\Delta t$ from their respective distributions. We check whether the resultant flux is within the defined flux range. If so, we move to simulate next burst. If not, we discard the $S_{i}/S_{i+1}$ value and draw another one. We repeat the process until the simulated flux is within the pre-defined flux range. This simulation effectively takes into account the flux distribution of FRB 121102. Indeed, after simulations of multiple bursts, our simulated burst flux distribution is consistent with the flux distribution of FRB 121102 (aside from the difference in the normalization factor). 

For each non-repeating FRB, we simulate the source based on the above procedure for a duration of the total observing time scale for the source. For each burst that is generated from the simulation, we compare it with the flux sensitivity threshold based on the S/N of the detected burst and the relative flux between the simulated burst and the initial detected burst. Only those bursts that are detectable by Parkes are kept. 

\begin{figure}
	\plotone{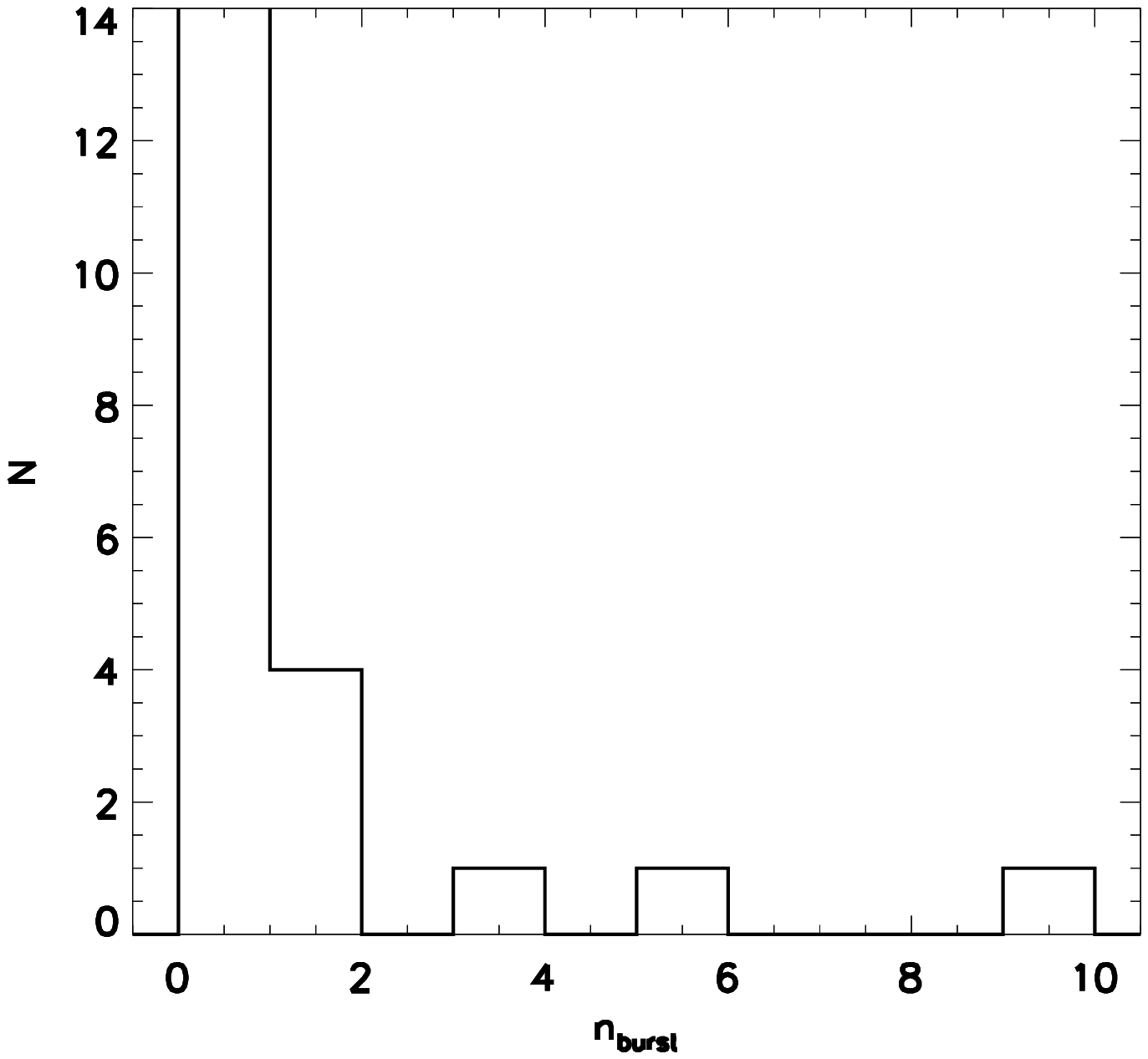}
	\plotone{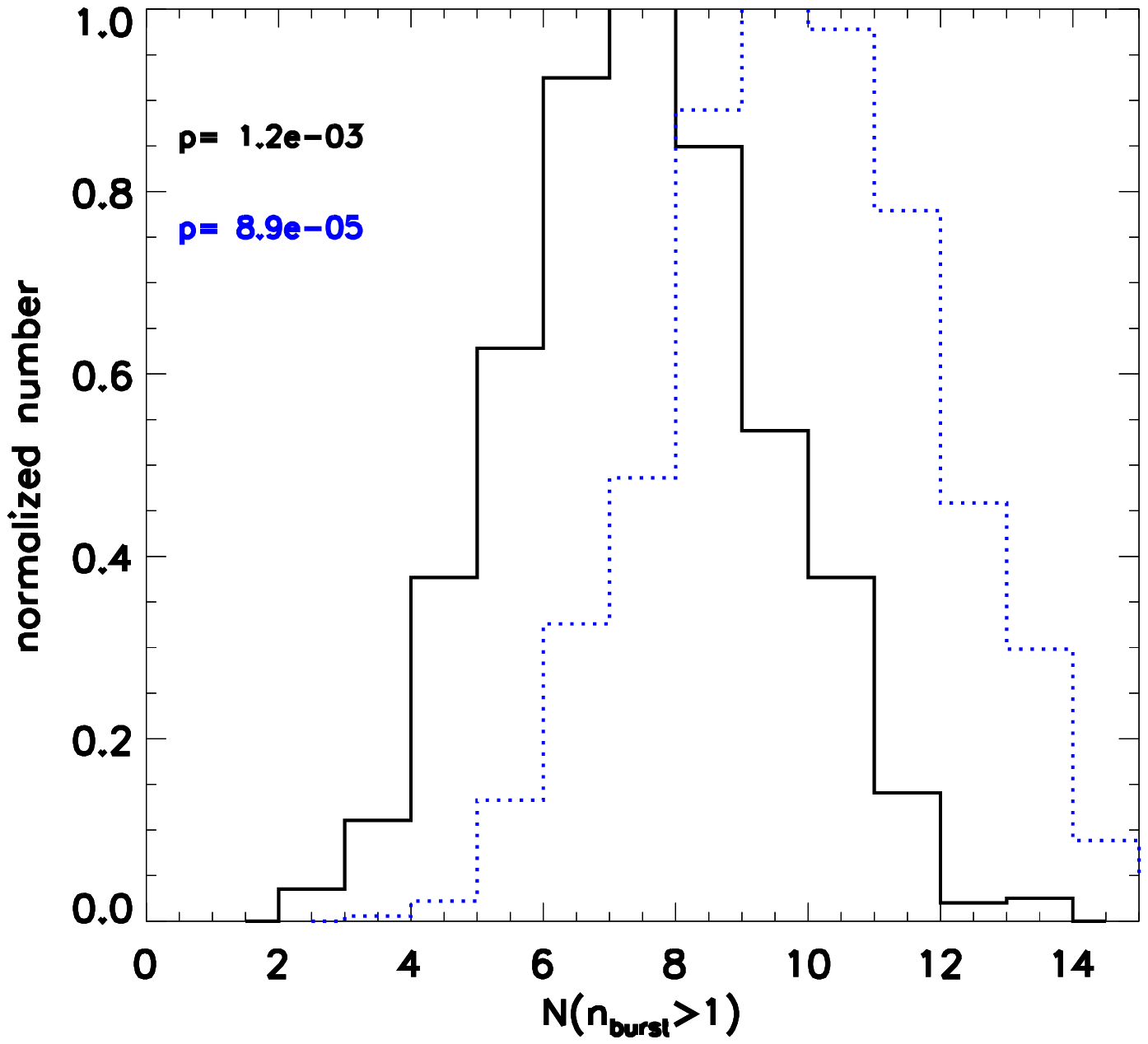}
	\caption{{\em Upper panel:} One realization of the simulation to show the histogram of number of bursts detectable bursts for the 21 FRB sources. {\em Lower panel:} The distributions of the total number of the detectable bursts after 10,000 simulations for Case I (black solid) and Case II (blue dotted), respectively. The corresponding non-detection probabilities are also marked.} 
	\label{fig:3}
\end{figure}

We focus on the most conservative Case I. Figure \ref{fig:3} upper panel shows one realization of the simulations, which is the histogram of the detected burst number besides the already detected one. One can see that for this realization, even though most sources (14 altogether) are consistent with having zero detected bursts (consistent with observations), other 7 sources all have extra detected bursts. One source even has 8 detectable bursts. This is inconsistent with the observations.

To quantify the probability of non-detection of any repeating burst from all 21 non-repeating FRBs, we perform 10,000 simulations and draw the probability distribution of the expected detectable bursts from all 21 sources. This is shown in the black histogram in Figure \ref{fig:3} lower panel. One can see the most probable case is to have 7 bursts detected from these sources. The probability of having no detection is $p \sim 1.2\times 10^{-3}$. If one considers the less conservative normalization (Case II) by assuming that the detected bursts are typical for FRBs, the peak distribution of the detectable bursts shifts to 9 (blue dotted histogram), and the non-detection probability is dropped to $p \sim  8.9 \times 10^{-5}$. 

We therefore conclude that the assumption that all FRBs are similar to FRB 121102 has a very low probability to be consistent with the data.  There is strong evidence of more than one population of FRBs.

\section{Conclusions and discussion}

Using the published data, we compare the repeating FRB 121102 and other FRBs in the $\Delta t - (S_i/S_{i+1})$ plane. We find that some bursts of the repeater are well separated from other FRBs in the plane. In particular, they occupy a region that has a much smaller duty cycle and a smaller flux ratio than other bursts. The conclusion remains valid if one only chooses the bursts that are detectable with the Parkes radio telescope. Dozens of hours of telescope time have been spent on the follow-up observations of other FRBs, yet none of them have been observed to repeat. Our Monte Carlo simulations suggest that the probability of non-detection of any burst in the 21 FRB sources is $\sim (10^{-4}-10^{-3})$. We therefore draw the conclusion that the repeating FRB 121102 is not representative of the entire FRB population, and that more than one population of FRBs is needed to account for the current data.

Since only the published data are used to perform this study, our conclusion is very conservative. Many more bursts have been detected from the repeating FRB 121102 (L. Spitler, 2017, http://aspen17.phys.wvu.edu/Spitler.pdf), and many more hours have been spent on other FRBs without a detection of repeating FRBs. This suggests that FRB 121102 and other FRBs should be even more separated in the $\Delta t - (S_i/S_{i+1})$ plane, which further strengthens our conclusion. 

One possibility is that the unusually active behavior of FRB 121102 is a result of a special amplification effect. For example, some propagation effects, such as scintillation-induced intermittency or plasma lensing \citep[e.g.][]{cordes97,cordes17}, may play a role of facilitating the detections of the bursts from FRB 121102, since its Galactic latitude is relatively low. Even though such a possibility is plausible, we notice that if other FRBs have a similar $S_i/S_{i+1}$ distribution as FRB 121102, still more than one burst should have been detected even if they were not amplified. This may be seen by artificially raising the detecting threshold for the FRB 121102 bursts (e.g. Fig. \ref{fig:ts2}).
We therefore suggest that the propagation effect cannot explain the dichotomy of the repeating behaviors of FRB 121102 and other FRBs.

The difference between FRB 121102 and other repeating FRBs is likely physical. One may consider two possibilities. First, maybe all FRBs share the same progenitor system, but FRB 121102 is unusually active in terms of producing repeating bursts. Within the spindown and magnetic powered models \citep[e.g.][]{Cordes2016,Connor2016,katz16,Metzger2017,Kashi2017,Cao2017,ZZ17}, a more active central engine should correspond to a younger age. According to this picture, other FRBs should have older ages than the repeater. However, the essentially zero evolution of DM for the repeater \citep{Spitler2016,Scholz2016} places a lower limit on the age of the putative supernova remnant associated with the pulsar \citep[e.g.][]{Metzger2017,Yang2017}. This is in apparent conflict with the putative FRB 110220/140514 association \citep{Piro2017}, which has a much more rapid DM evolution but less frequent bursts. 

A more likely possibility is that there are more than one population of FRBs. The repeating FRB 121102 has a distinct progenitor system than (at least some of) other FRBs. The current data already showed strong evidence of such a possibility. It is possible that (at least some of) other FRBs are produced in catastrophic events so that they are intrinsically non-repeating FRBs. Blitzars \citep{FR2014, Zhang2014} and merger of compact stars \citep{Tot2013, Kashi2013, Zhang2016, wang16} remain attractive candidate progenitors for these non-repeating FRBs. A future coincident detection of an FRB with a catastrophic event (e.g. a gravitational wave chirp signal) would unambiguously establish such a non-repeating FRB population.

\acknowledgments 
We thank an anonymous referee for very helpful suggestions. This work is partially supported by NASA through grants NNX15AK85G and NNX14AF85G to UNLV.

\end{document}